\documentclass[12pt,preprint]{aastex}

\newcommand{\etal}{{et\thinspace al.} }
\def\spose#1{\hbox to 0pt{#1\hss}}
\def\simlt{\mathrel{\spose{\lower 3pt\hbox{$\mathchar"218$}}
     \raise 2.0pt\hbox{$\mathchar"13C$}}}
\def\simgt{\mathrel{\spose{\lower 3pt\hbox{$\mathchar"218$}}
     \raise 2.0pt\hbox{$\mathchar"13E$}}}
\def\simpropto{\mathrel{\spose{\lower 3pt\hbox{$\mathchar"218$}}
     \raise 2.0pt\hbox{$\propto$}}}

\begin{document}

\title{The Infrared Counterparts of the Optically Unidentified CDF-S 1Ms
Sources\altaffilmark{1}}
\author{Haojing Yan\altaffilmark{2},
Rogier A. Windhorst\altaffilmark{2},
Huub J. A. R\"{o}ttgering\altaffilmark{3},
Seth H. Cohen\altaffilmark{2},
Stephen C. Odewahn\altaffilmark{2},
Scott C. Chapman\altaffilmark{4},
and William C. Keel\altaffilmark{5}
}

\altaffiltext{1}{Based on observations made with the European Southern
Observatory telescopes obtained from the ESO/ST-ECF Science Archive Facility.}
\altaffiltext{2}{Department of Physics and Astronomy, Arizona State University,
Tempe, AZ 85287-1504;  Haojing.Yan, Rogier.Windhorst, Seth.Cohen, Stephen.Odewahn@asu.edu}
\altaffiltext{3}{Leiden Observatory, P. O. Box 9513, 2300 RA Leiden, The Netherlands; rottgeri@strw.leidenuniv.nl}
\altaffiltext{4}{Department of Physics, California Institute of Technology,
1200 E. California Blvd., Pasadena, CA 91125; schapman@irastro.caltech.edu}
\altaffiltext{5}{Department of Physics and Astronomy, University of Alabama,
       Tuscaloosa, AL 35487; keel@bildad.astr.ua.edu}

\begin{abstract}

  The Chandra Deep Field South (CDF-S) 1Ms exposure produced a catalog of 346
X-ray sources, of which 59 were not visible on the VLT/FORS1 and the 
ESO-MPI/WFI deep $R$-band images to a limit of $R_{vega}$=26.1--26.7 mag.
Using the first release of the ESO VLT/ISAAC $JHK_s$ data on the CDF-S, we
identified six of the twelve such objects that were within the coverage of
these IR observations. The VLT/FORS1 $I$-band data further confirms that five
of these six objects are undetected in the optical.  The
photometric properties of these six counterparts are compared against those of
the optically brighter counterparts of Chandra sources in the same
field. We found that the location of these optically brighter Chandra sources
in the near-IR color space was bifurcated, with the color of one branch
being consistent with that of E/S0 galaxies at $0\leq z\leq 1.5$ and the other 
branch being
consistent with that of unreddened AGN/QSOs at $0\leq z\leq 3.5$. The six
counterparts that we identified seemed to lie on the E/S0 branch and its
extension, suggesting that these X-ray source hosts are mostly luminous E/S0
galaxies ($M_V\sim -20$ mag in AB
system) at $1\leq z\leq 2.5$. On the other hand, some of them can
also be explained by AGN/QSOs over a wide redshift range ($0\leq z \leq 5$) if
a range of internal extinction ($A_V=0$--1 mag) is allowed. However,
the later interpretation requires fine-tuning extinction together with
redshift for these objects individually. If they are indeed AGN/QSOs, the
most luminous of them is just barely qualified for being a QSO. Finally, we
point out that neither high-redshift ($z > 5$) star-forming galaxies nor
irregular galaxies at lower redshift can be a viable explanation to the nature
of these six counterparts.

\end{abstract}


\section{Introduction}

   The investigations conducted prior to the launches of Chandra and XMM
have speculated that a significant fraction of the X-ray background (XRB) is
produced by discrete sources, predominantly AGNs (e.g. Hasinger \etal 1998).
With its superb spatial resolution, Chandra resolved nearly
100\% of the X-ray background in the 0.5--8.0 keV regime into discrete sources
(e.g. Mushotzsky \etal 2000), and thus changed the theme of the XRB study into
investigating the nature of the discrete X-ray sources and their constraints
on galaxy evolution. With the active optical and infrared follow-ups
around the two deepest Chandra fields, Chandra Deep Field
North (CDF-N; Brandt \etal 2001b) and South (CDF-S; Giacconi \etal 2002),
we have gained extensive knowledge about the faint X-ray source population.
It is now generally believed that the majority of these sources are 
dust-obscured type-II AGNs at $z$=0--2, while a smaller fraction
of them are classic type-I AGNs out to $z\simeq 4$ (e.g., Alexander \etal 2001;
Tozzi \etal 2001; Norman \etal 2002; Rosati \etal 2002).

   However, a significant number of these X-ray sources are still
optically unidentified and thus their nature remains uncertain.
Within an 8'.6$\times$8'.6 area at the center of the CDF-N,
the multi-wavelength investigation of 
Alexander \etal (2001) indicates that fifteen, or about 10\% of the total X-ray
sources do not have counterparts to a 2\,$\sigma$ limit of $I$ = 25.3
mag (all magnitudes quoted in this letter are in the Vega system unless
otherwise noted).
Five out of these fifteen sources, however, have near-IR counterparts on
the $HK^{\prime}$-band images of Barger \etal (1999). Similarly, Giacconi \etal 
(2002) concluded that 59, or about 17\% of the total X-ray sources in the CDF-S
were not optically identified at a limit of $R$ = 26.0--26.7 mag. These facts
raise a very natural question: are these optically invisible sources merely
the fainter tail of the already identified population, or do they constitute a
separate population of their own?

   In this letter, we study the nature of the optically unidentified sources
in Giacconi \etal (2002), which we shall refer to as "$R$-unidentified sources".
We utilize the deep VLT/ISAAC $JHK_s$ and VLT/FORS1
$I$-band data, both of which are now publicly available from the ESO science
archive.  The analysis of these data are described in \S 2, and the IR and 
$I$-band identifications are shown in \S 3, where we presnet IR identifications
of six such X-ray sources. In \S 4 we explore
the possible nature of these sources based on their optical to near-IR 
photometric properties.  We conclude with a summary in \S 5.

\section{Data Analysis}

   As part of the supporting observations of the {\it Great Observatories
Origins Deep Survey} (GOODS; Dickinson \& Giavalisco 2002) fields, the ESO
VLT/ISAAC has imaged a large portion of the CDF-S in $J$, $H$, and $K_s$-bands
(hereafter $JHK_s$).  The first release of these fully processed and stacked
data consists of six continuous fields, and covered about 50 $arcmin^2$ at the
center of the CDF-S.
To further constrain the optical characteristics of the Chandra sources, 
we also independently reduced the whole set of CDF-S $I$-band VLT/FORS1
images (3 hours of exposure, ESO Prog. ID. 64.O-0621(A), PI Gilmozzi)
retrieved from the ESO science archive. These $I$-band data, consisting of
four slightly overlapped fields, extend $\sim$ 150 $arcmin^2$ and fully
cover the area that the VLT/ISAAC has imaged in the $JHK_s$.
The photometry on both the $JHK_s$ data and the $I$-band data was done by using
the SExtractor routine of Bertin \& Arnouts (1996).

   For each ISAAC field, the $JHK_s$ images were combined into a master stack, 
which was then used for extracting sources and defining apertures. While the
photometry was performed on the individual passbands, a same aperture defined
from the master stack was used for a given source in each passband (i.e., 
``matched-aperture photometry"). For source detection, we used a $5\times 5$
Gaussian smoothing kernel with the FWHM of 2.5 pixels, which is approximately
the same as the FWHM of a point source PSF ($\sim 0.5''$) on the master stack.
We set 1.5 $\sigma$ as the detecting threshold, and required at least 5 
connected pixels above the threshold for a source to be detected. As the vast
majority of the sources are extended, we used total magnitude (corresponding
to the {\it mag-auto} option in SExtractor) for the photometry on the individual
passbands. The zeropoints recorded in the image headers were used to convert
counts into Vega magnitudes. 

   We assessed the $JHK_s$ limits in the following way. The representative 
error reported by SExtractor was used to calculate the S/N of each extracted
source, assuming the reported magnitude error ($\Delta m$) and the S/N follow
the simple relation as $\Delta m = 1.0857/(S/N)$. Only the sources with 
$S/N \geq 3\,\sigma$ were included in the assessment. The source count
histogram was constructed for each band, and the magnitudes at which the counts
dropped to 50\% of the peak values were defined as the survey limits.
These numbers are: $J$=24.2, $H$=23.4 and $K_s$=23.2 mag.

   Source detection and photometry on the $I$-band stacks were done in the
same manner as in analyzing the ISAAC data excepted that the detection 
threshold was lowered to $1 \sigma$ to include as faint sources as possible. As the ESO Imaging Survey
(EIS; Arnouts \etal 2002) covered the entire CDF-S, we used their $I$-band 
source catalog to calibrated our photometry. We first removed the Galactic
extinction ($A_I$=0.02 mag) that they applied to their results, and converted
the magnitudes from AB system, which they used, to Vega system by using
$I_{vega}=I_{AB}-0.48$.  The survey limits, derived in the same way as in the
ISAAC data analysis, are approximately 25.1 -- 25.2 mag accounting for
field-to-field variation.

\section {Source Cross-matching}

   Out of the 346 Chandra sources, 77 have their locations falling within the
$JHK_s$ coverage, among which twelve are $R$-unidentified sources.
We found that six of these twelve objects could be matched with the $JHK_s$
sources within a matching radius of 1.5", all of which were strikingly
significant detections (at 3.5--9.0\,$\sigma$ level) in the $K_s$-band. After 
visually examining these matched sources on screen and comparing against 
the $R$-band image stamps given in Giacconi \etal (2002), we concluded that
these six IR counterparts were real identifications. 

   We then cross-matched the $R$-unidentified sources with our $I$-band
images. While 35 of them have their locations within the $I$-band coverage, only
one counterpart was found. This fact further suggests that most of these 
objects are
indeed optically faint. However, the only counterpart, which is also one of the
six $JHK_s$ sources, was detected at above $7\,\sigma$ level.

   Table 1 lists the positions and the photometric characteristics of the six
objects that we identified, together with their X-ray hardness ratios ($HR$)
taken from Giacconi \etal (2002). These six sources are of moderate X-ray
fluxes in 0.5--2 kev, with two of them falling toward the faint end
($\sim 8.5\times 10^{-17}$ erg/s/cm$^2$).  We note that they tend to have 
negative X-ray hardness ratios. In fact, three of them were detected in the
soft channel only ($HR=-1$). Assuming that their underneath X-ray sources are
AGNs, it is possible to determine their types based on their $HR$ values. If
we use $HR\simeq -0.1$ as the dividing line between low X-ray luminosity
($L_X$) type-II AGN and high $L_X$ type-I AGN (e.g. Rosati \etal 2002), three
of these six Chandra sources are type-I AGNs while the remaining two are
type-II AGNs. However, such a division should be used with caution here,
as our case is of small number statistics (cf. Fig. 3 of Rosati \etal 2002).
The excellent image quality (point source FWHM $\sim 0.5"$ in both IR and 
$I$-band) allowed us to at least determine if these sources are of extended
morphology. Based on the $K_s$-band image, where the S/N of the objects
is the highest, we determined that three of them were extended sources, and
two of them were point sources. The remaining one is hard to tell because it
is close to the $K_s$-band field edge and its background is much noisier than
that of the others.

\section {Nature of the Six IR Counterparts}

\subsection {Possible Candidates}

   Although these $R$-unidentified sources could simply be the fainter tail
of the brighter objects that have been optically identified, we should not
limit our investigation to only this possibility.
For this reason, we should consider a wide variety of
candidates when exploring the nature of these X-ray source hosts.

   As Alexander \etal (2001) argued, it is very unlikely that the objects in
the Galaxy could be a major contribution to optically faint X-ray sources.
This possibility is even smaller for the CDF-S because it is at high galactic
latitude. Thus we consider four types of extragalactic objects as the 
primary candidates, namely, elliptical galaxies (E/S0), AGNs/QSOs (AGN), 
irregular galaxies (Irr), and very young galaxies at high redshift
($z>5$; High-z). We do not include
spiral galaxies for the sake of simplicity, as their photometric properties
in the $I$-band and the near-IR lie in between the ellipticals and the
irregulars.  We compare the six IR counterparts against these four types of
candidate ins the $IJHK_s$ color space. We consider the redshift range of 
$5 \leq z \leq 10.5$ for high redshift young galaxies, and $0\leq z \leq 5$
for all the other three types of candidates.  
We also consider the effect of internal dust extinction
of $A_V$=0.1--3 mag, following the standard Milky Way extinction law. For the
wavelength range shortward of 2000${\rm \AA}$, we use the average values of 
the LMC, SMC and the Galaxy (e.g. Calzetti, Kinney \& Storchi-Bergmann 1994).

\subsection {Color-Color Diagram Diagnosis}

   Because five of these six objects are only visible in the three IR passbands,
it is not realistic to perform rigorous photometric redshift analysis based on
such limited information. At this stage, we chose to use color-color diagrams
as our primary diagnostic tools, which are less quantitative but more robust. 
The invisibility of these sources in the $R$-band and blueward was used as a
consistency check for each of the four possibilities, and was also used as one
of the inputs (detailed below) to
constrain the rest-frame luminosity of the sources.

   We simulated the colors of the candidate objects by using their
representative spectral templates. The templates for E/S0 and Irr were taken
from Coleman, Wu \& Weedman (1980) with the extension to UV and IR according
to Bruzual \& Charlot (1993), and the one for AGN was taken from 
Vanden Berk \etal (2002; using $f_{\lambda} \propto \lambda ^{-0.45}$ to extend
its continuum to the near-IR). For young, high redshift galaxies, we used a 0.1
Gyr age, solar metalicity template from Bruzual \& Charlot (1993). These 
rest-frame spectral templates were redshifted in $\Delta z$=0.5 step, and
attenuated for the cosmic H I absorption according to Madau (1994). The 
resulting spectra were then convolved with the system responses, mainly the
filter transmission and the detector quantum efficiency, to generate the
colors in the AB magnitude system. When considering dust extinction, the colors
were adjusted according to the rest-frame spectral range sampled by the 
relevant passbands. The colors were finally transfered into the Vega system by
adopting the appropriate zeropoints (e.g. Waddington \etal 1999).

   Fig. 1 shows the $(H-K_s)$ vs. $(J-K_s)$ color-color diagrams in four 
separate panels for the four cases. The field objects with $S/N \geq 3$
in all three bands are shown as the green dots, and the Chandra sources
that have been identified in $R$-band are shown as the blue squares.
The six IR counterparts are plotted with
different symbols based on their hardness ratios: the three solid triangles 
are the ones with $HR=-1$, the two open hexagons that have $-1< HR < 0.1$,
and the open square is the source that has $HR > 0.1$.
Their error bars indicate the photometric errors in their colors, which
are calculated as the square-root of the quadratic sum of their photometric errors
in the corresponding passbands.
The unreddened and reddened colors of the simulated objects at different 
redshifts are plotted in black crosses and solid red squares, respectively
(in the case of E/S0, only unreddened colors are shown). For clarity, only
the optimal results are plotted for the reddened colors. These optimal
reddening values are $A_V=1.6, 1.0$ and 1.6 mag for Irr, AGN and High-z cases,
respectively.
A couple of redshifts are marked to indicate the direction along which
the color of the simulate objects changes with redshift. 

   Similarly, Fig. 2 shows the $(H-K_s)$ vs. $(I-K_s)$ color-color diagrams.
The High-z case is not included in this figure, because galaxies at $z >$ 5 
will {\it drop-out} from the $I$-band and thus they cannot put meaningful 
$(I-K_s)$ constraint on those IR counterparts that are also $I$-band
drop-outs.  All the legends are the same as in Fig.1, except that the five
of the six IR counterparts that were not detected in the $I$-band are not
plotted with error-bars. Instead, we put upward arrows on them to indicate
their $(I-K_s)$ color limits.

   It is intriguing to note that the location of those optically visible 
Chandra sources in the IR color space is bifurcated, as can be clearly seen in
Fig. 1. The upper branch coincides with the E/S0 track from 
$0\leq z\leq 1.5$ while the lower branch seems to be consistent with the AGN 
track from $0\leq z\leq 3.5$. This interpretation does not conflict with the
investigation on the brighter sources in Rosati \etal (2002; their Fig. 5.).
The six objects that we identified lie
on the upper branch and its extension following the E/S0 track up to 
$z\simeq 2.5$ (see below).

\subsection {Comments on Individual Cases}

   Using the two color-color diagrams and the source brightness, we are able
to set stringent constraints on the nature of these six IR counterparts. We
comment on the above-mentioned four possibilities individually.

   1. $E/S0$\,\, From Fig. 1 and 2 it is obvious that the six sources have
colors consistent with unreddened E/S0 galaxies at $z$=0.5--2.5, and
introducing internal extinction does not improve the fit. Considering
the brightness of these objects, this interpretation is also consistent with
their invisibility in the $R$-band and
blueward.  While No. 593 (the lowest triangle in Fig. 1), which was
detected in the $I$-band as well, is probably at $z$=0.5, the other five 
objects are likely at $1.0 \leq z \leq 2.5$. If we use $z$=2.0 as their
representative redshift, their averaged rest-frame $V$-band absolute magnitude
is $\sim M_V$ = --20.1 mag (in AB system), indicating that they are giant 
ellipticals. In contrast, No. 593 has only $M_V \simeq -16.7$ mag if it is at
$z$=0.5, falling in the dwarf elliptical regime.

   2. $AGN$\,\, While they can reasonably explain the colors of the optically
bright Chandra sources, unreddened AGNs at $0\leq z \leq 5$ cannot produce 
colors similar to the six sources (see especially Fig. 2). If internal 
reddening is introduced, the matching will be improved. However, this
interpretation involves both the extinction and the redshift as free
parameters, and requires both of them be adjusted for each source individually.
If the extinction is forced to be fixed, the best fit is obtained with
$A_V=1.0mag$, which allows three of the six source
(open symbols) to be explained, i.e., at 
$z\simeq$ 0--1 or $z\simeq$ 4--5. If we take the first interpretation for 
these $three$ objects, their absence from $I$-band and blueward means
they can only be at $z\simeq$ 0.8--1.0 to have a reasonable intrinsic luminosity
that still qualified for AGNs ($M$(3800--4200\,{\rm $\AA$}) $\sim -18$ to $-19$
mag after de-reddening). If the later interpretation is real, on the other
hand, their absolute magnitudes make them fall in the low-luminosity end of
QSOs ($M$(3800--4200\,{\rm $\AA$}) $\sim -23$ to $-24$ mag after de-reddening).
The cumulative QSO surface density inferred from this interpretation (2--3 in
50 $arcmin^2$) is consistent with the QSO luminosity function extrapolated
from that is known at lower redshifts (e.g. Boyle \etal 2000). Particularly,
we noted that No. 593 did not fit in this interpretation no matter how the
extinction and the redshift were adjusted.

   3. $Irr$\,\, Similar to the AGN case, the colors of unreddened irregular
galaxies match the optically bright Chandra sources but cannot match those
of the six $R$-unidentified sources. Introducing internal reddening cannot
solve the problem, as Fig. 2 shows. Thus irregular galaxies are not likely
a viable explanation to the nature of the six sources.

   4. $High-z$\,\, The colors of unreddened high-z galaxies are not consistent 
with the six sources. Fine-tunning internal reddening value can produce a
resonable fit either at the redshift range $7.5\leq z\leq 8.5$ or at 
$5.0\leq z \leq 5.2$.
The reddening value thus required is $A_V\simeq 1.0$ mag. However, in either
case, the cumulative surface density of such galaxies are at least one order
of magnitude higher than a reasonable luminosity function would predict 
(following the same methodology as in Yan \etal 2002). Thus we conclude that
this hypothesis is not a viable interpretation either.

\section{Summary}

   We present deep optical and near-IR identifications of the $R$-unidentified
X-ray sources in the CDF-S, using the deep VLT/ISAAC and FORS1 imaging
data publicly available from the ESO science archive. Alexander \etal
(2001, 2002) did a similar work over a comparable size of area around the
CDF-N, who largely focused on the optically faint {\it but visible
sources}. The VLT data that we used reached at least two magnitudes deeper than theirs
in the near-IR, and was at least twice as deep in the $I$-band. More 
importantly, the three near-IR bands used in this study allow us to further
constrain the nature of such sources.

   We found six IR counterparts for the CDF-S 1Ms sources that were not 
optically identified to $R$=26.1--26.7 mag, making it possible to carry out
spectroscopic study in the future. The X-ray hardness ratio of these six
Chandra sources are more negative, opposite to the trend found among the 
similar, but optically brighter sources in the CDF-N (Alexander \etal 2001).
The multi-band photometric properties of these six sources are compared against
those of the optically visible Chandra sources in the same field.
We found that the location of these optically brighter Chandra sources
in the near-IR color space was bifurcated, with the color of one branch
being consistent with that of E/S0 galaxies at $0\leq z\leq 1.5$ and the other
branch being consistent with that of unreddened AGN/QSOs at $0\leq z\leq 3.5$.
The six counterparts that we identified seem to lie on the E/S0 branch and 
its extension, suggesting that these X-ray source hosts are mostly luminous
E/S0 galaxies ($M_V\sim -20$ mag in AB system) at $1\leq z\leq 2.5$.
Such luminous ellipticals that host X-ray sources might be similar to those
found by Cowie \etal (2001), which are amplified by a foreground cluster. 
 On the other hand, some of these six sources can
also be explained by AGN/QSOs over a wide redshift range ($0\leq z \leq 5$) if
various internal extinction ($A_V=0$--1 mag) is allowed. However,
the later interpretation requires fine-tuning extinction together with
redshift for these objects individually. If they are indeed AGN/QSOs, the
most luminous of them is just barely qualified for being a QSO. Finally, we
point out that neither high-redshift ($z > 5$) star-forming galaxies nor
irregular galaxies at lower redshift can be a viable explanation to the nature
of these six counterparts.

\acknowledgments

\begin{table}
\caption{}
\begin{center}
\begin{tabular}{ccccccccc}\tableline \tableline

ID\tablenotemark{a}& RA \& DEC (J2000)\tablenotemark{b}& $I$\tablenotemark{c}& $J$\tablenotemark{d}& $H$\tablenotemark{d}& $K_s$\tablenotemark{d}& P/E \tablenotemark{e}& HR\tablenotemark{f} \\ \tableline
 79 & 3:32:38.04   -27:46:26.2 &   $>25.1$      &  23.21$\pm$0.23 &  21.88$\pm$0.17 &  20.95$\pm$0.12 & P   & -0.42 \\
593 & 3:32:14.79   -27:44:02.5 &  24.86$\pm$0.18 &  22.97$\pm$0.21 &  21.66$\pm$0.14 &  21.18$\pm$0.13 & E   & -1.00 \\
221 & 3:32:08.91   -27:44:24.8 &   $>25.1$         &  25.08$\pm$0.66 &  22.41$\pm$0.21 &  21.83$\pm$0.17 & P   & -1.00 \\
515 & 3:32:32.17   -27:46:51.5 &   $>25.1$         &  24.68$\pm$0.76 &  22.91$\pm$0.27 &  21.90$\pm$0.21 & E   &  0.41 \\
561 & 3:32:22.44   -27:45:43.8 &   $>25.1$         &  25.73$\pm$0.86 &  23.49$\pm$0.38 &  22.68$\pm$0.29 & E   & -1.00 \\
201 & 3:32:39.06   -27:44:39.3 &   $>25.1$         &  24.79$\pm$0.58 &  22.62$\pm$0.21 &  21.52$\pm$0.16 & ... & -0.06 \\

\tableline
\end{tabular}
\end{center}
\tablenotetext{a}{Source ID as in Giacconi \etal (2002; their XID).}
\tablenotetext{b}{RA and DEC as measured on the $JHK_s$ images.}
\tablenotetext{c}{The "auto" magnitude (Kron magnitude) as derived by SExtractor, in Vega system.}
\tablenotetext{d}{Same as c; aperture used for each source is the same in all these three bands.}
\tablenotetext{e}{Point (P) or extended (E) source, based on the $K_s$-band 
images, on which a point source has a FWHM of only $\sim 0.5"$.}
\tablenotetext{f}{X-ray hardness ratio ($HR$), defined as (H-S)/(H+S), where H 
and S are the counts in the hard and soft channel, respectively. Values are 
taken from Giacconi \etal (2002).}

\end{table}

\clearpage 

\begin{figure}
\plotone{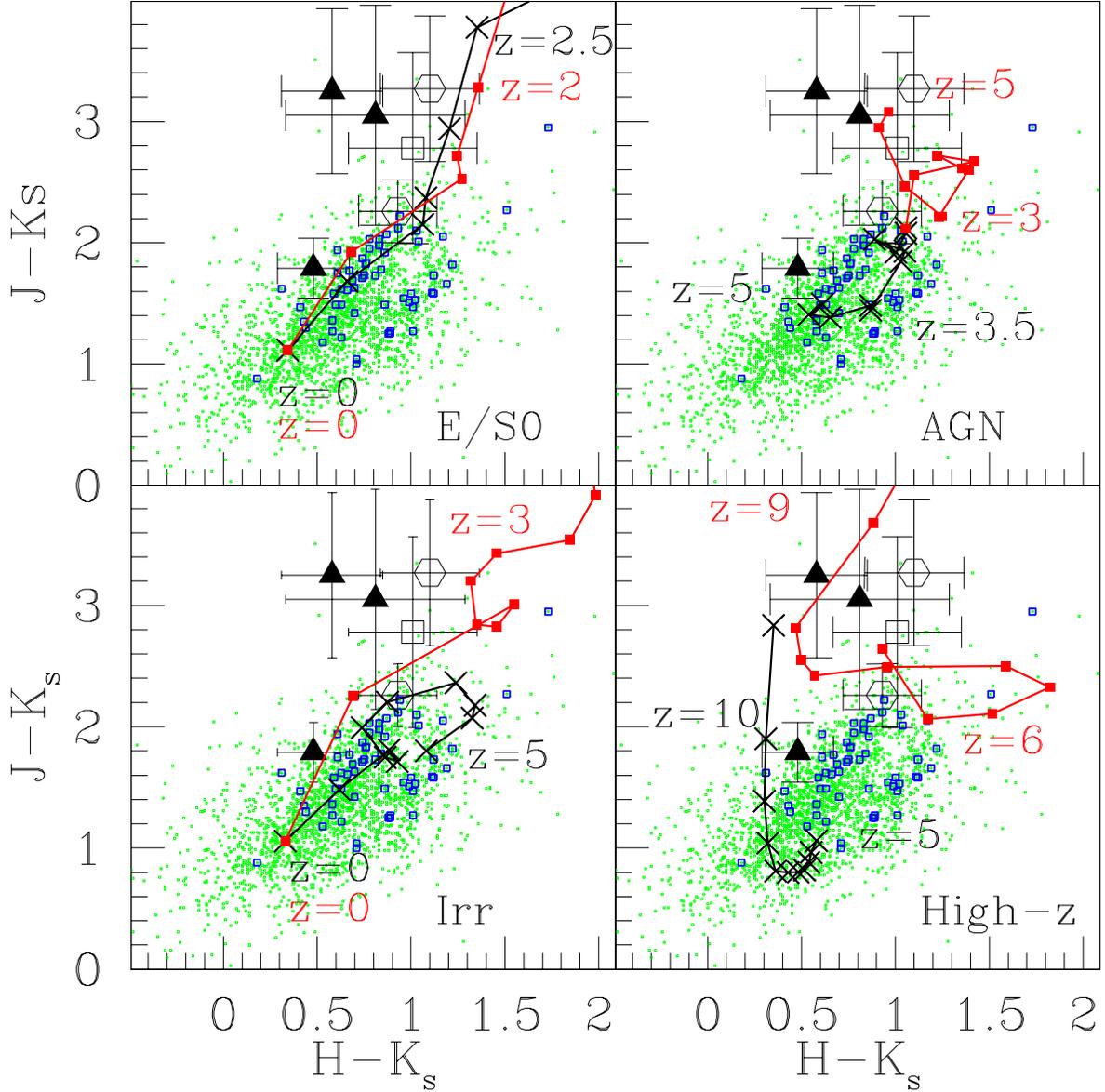}
\caption{($H-K_s$) vs. ($J-K_s$) color-color diagram as the diagnostic tool
of the source nature. The green dots are the field objects. The open blue 
squares are the optically bright Chandra sources, whose locus shows clear
bifurcated structure.
The large symbols with error bars
are the six $R$-unidentified Chandra sources that were identified in the IR:
the three solid triangles are the sources with HR=-1,
the two open hexagons are the ones with HR$<$-0.1, and the open square is the
source d with that has HR$>$-0.1.  Their colors
are compared against the simulated colors of four types of candidate in
separate panels. The redshift ranges considered are $0\leq z\leq 5$ for E/S0,
Irr and AGN/QSO, and $5\leq z\leq 10.5$ for High-z. Dust extinction ranging 
from $A_V$=0.5 to 3.0 mag is also considered. For
clarity only the cases with the optimal amount of dust extinction are shown
($A_V=0.5$ mag for E/S0, $A_V=$1.0 mag for AGN/QSO and High-z, and $A_V=$1.6 mag for Irr). The 
unreddened colors of these simulated objects are shown as black crosses, while
their reddened colors are show as filled red squares. A couple of redshifts
are also marked (in both unreddened and reddened situations) to indicate the
direction on which the colors change with the redshifts. 
}
\end{figure}

\begin{figure}
\plotone{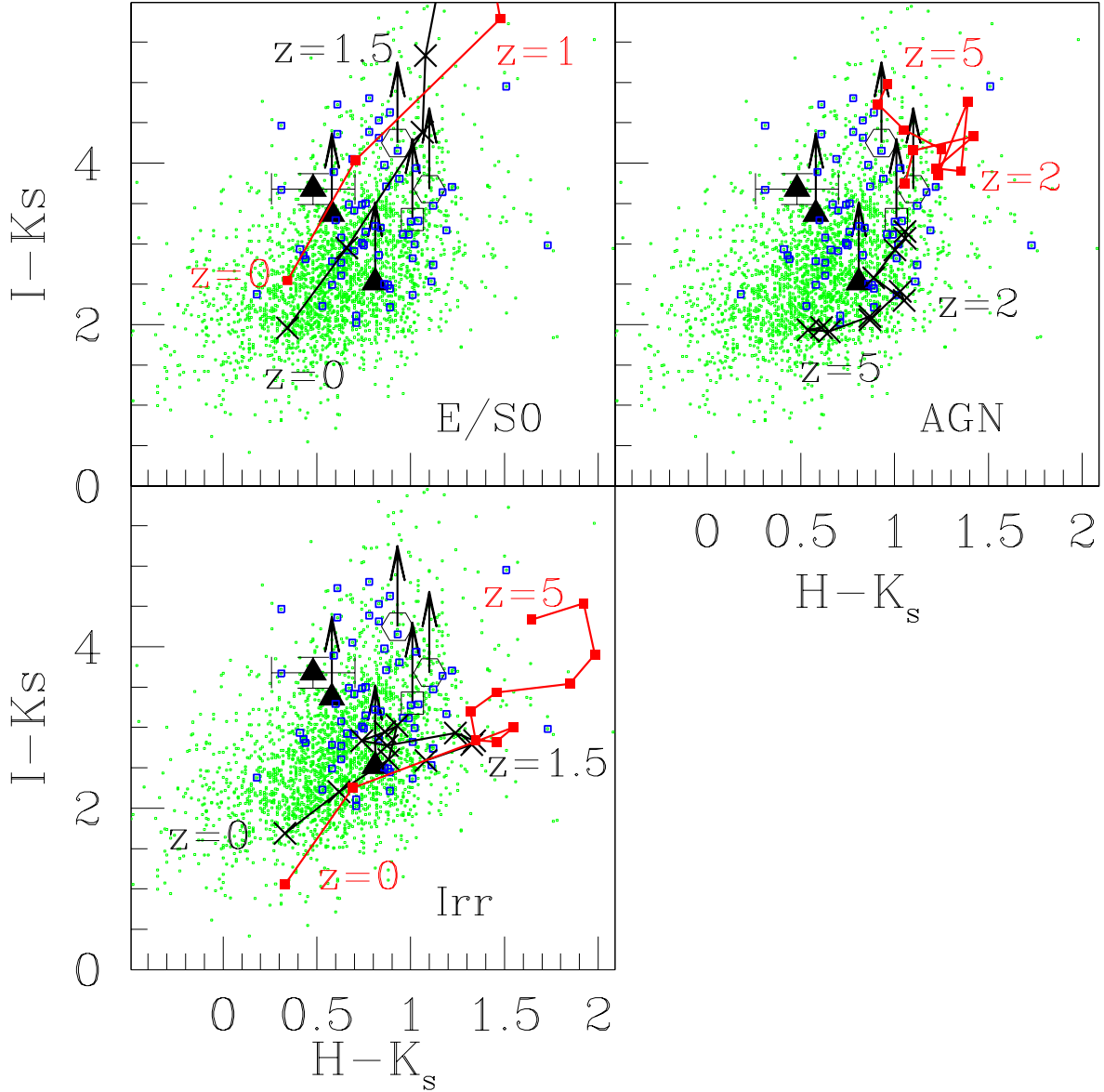}
\caption{Same as Fig. 1, but for the $(I-K_s)$ vs. $(H-K_s)$ color-color 
diagram. The High-z case is not shown, as it cannot put further
constraint on the $(I-K_s)$ of the six IR counterparts (see text). The arrows
indicate the lower limit of the $(I-K_s)$ color for the five objects that
were not detected in the $I$-band. Very importantly, the case of irregular
galaxies does not pass this test because the $(I-K_s)$ of such objects are not
consistent with the colors of the IR counterparts, even considering the effect
of dust extinction.
}
\end{figure}

\end{document}